\documentclass{iau} 
\usepackage{graphicx}
\usepackage{natbib}
\newcommand\aj{{AJ}}%
\newcommand\apj{{ApJ}}%
\newcommand\aap{{A\&A}}%
\newcommand\mnras{{MNRAS}}%
\newcommand\pasp{{PASP}}%
\title[Radial mixing and Galactic thick and thin disks] 
{Impacts of Radial Mixing on the Galactic Thick and Thin Disks}

\author[Daisuke Kawata]   
{Daisuke Kawata$^1$}

\affiliation{$^1$Mullard Space Science Laboratory, University College London, Holmbury St. Mary, Dorking, Surrey, RH5 6NT, UK \\ email:{\tt d.kawata@ucl.ac.uk}}

\pubyear{2017}
\volume{334}  
\jname{Rediscovering our Galaxy}
\editors{C. Chiappini, I. Minchev, E.  Starkenburg \& M. Valentini, eds.}
\begin{document}

\maketitle

\begin{abstract}
Using N-body simulations of the Galactic disks, we qualitatively study how the metallicity distributions of the thick and thin disk stars are modified by radial mixing induced by the bar and spiral arms.  We show that radial mixing drives a positive vertical metallicity gradient in the mono-age disk population whose initial scale-height is constant and initial radial metallicity gradient is tight and negative. On the other hand, if the initial disk is flaring, with scale-height increasing with galactocentric radius, radial mixing leads to a negative vertical metallicity gradient, which is consistent with the current observed trend. We also discuss impacts of radial mixing on the metallicity distribution of the thick disk stars. By matching the metallicity distribution of N-body models to the SDSS/APOGEE data, we argue that the progenitor of the Milky Way's thick disk should not have a steep negative metallicity gradient.
\end{abstract}

\firstsection 
\section{Introduction}

 The evolution of radial metallicity gradients should provide strong constraints on disk formation scenarios \citep[e.g.][]{gpbsb13}, and hence it is important to compare gradients of different age populations of the Milky Way with Milky Way-like disk galaxies at different redshifts. However, it is not straightforward to infer initial metallicity distribution for mono-age populations at the time they formed, from the current Galactic metallicity distribution as a function of age, because radial mixing can alter the distributions \citep[e.g.][]{sb09a,mcm14,gkc15}. We refer to ``radial mixing" to describe the overall radial re-distribution due to both ``churning" and ``blurring" \citep{sb09a}. Churning indicates the change of angular momentum of stars, and blurring describes the radial re-distribution of stars due to their epicyclic motion. Churning can be split into two mechanisms, ``radial scattering" and ``co-rotation radial migration". Radial scattering describes the change of angular momentum with kinematic heating, i.e. the increase in random energy of the orbit. On the other hand, co-rotation radial migration indicates a gain or loss of angular momentum of stars at the co-rotation resonance of the spiral arms \citep{jsjb02} or bar, and does not involve significant kinematic heating. Here, we discuss the effect of radial mixing, without distinguishing these mechanisms.

Using simple numerical experiments based on N-body simulations of an idealised isolated galactic disk, we qualitatively study the impact of the radial mixing on the metallicity distributions of the thin and thick disk populations. Here, we consider chemically defined thick (high-[$\alpha$/Fe]) and thin (high-[$\alpha$/Fe]) disk, and also consider that thick disk population is generally older than thin disk population.

\begin{figure}
\begin{center}
\includegraphics[width=\hsize]{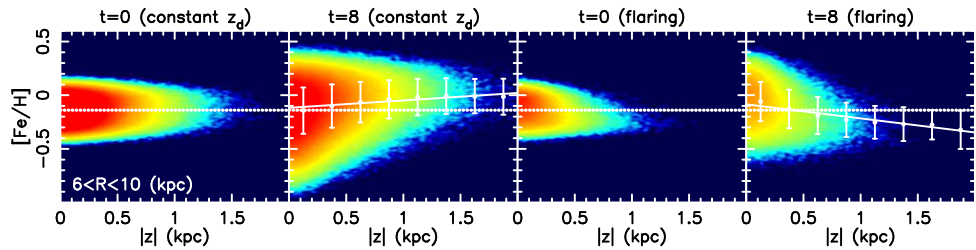} 
 \caption{Vertical metallicity distributions of the thin disk component for constant scale-height, $z_d$, model (left two panels) and flaring model (right two panels) in the radial ranges of $6<R<10$~kpc at $t=0$ and $8$ Gyr. The white dots and the error bars at $t=8$ show the mean and dispersion of [Fe/H] at different height, $|z|$, and the solid line shows the best fit straight line. The dotted line shows the vertical metallicity gradient assumed at $R=8$~kpc at $t=0$.}
\label{fig:zmet}
\end{center}
\end{figure}

\section{Vertical Metallicity Gradient and Flaring Disk} 
\label{sec:vZflaringthin}

Combining Gaia DR1 TGAS data \citep{gcbvpdbm16,llbhkhb16} and RAVE DR5 \citep{kkszmce17}, \citet{cklcsc17} analysed the vertical metallicity gradient of solar neighbourhood dwarf stars, and found that the older mono-age population shows a steeper negative vertical metallicity gradient \citep[see also][]{xlyhw15}. On the other hand, the thin disk population stars are expected to form within very thin high density molecular gas. It is a natural expectation that the star forming gas disk has almost no vertical metallicity gradient, because metal mixing in vertical direction should be quite efficient. Then, how was the observed negative vertical metallicity gradient of mono-age thin disk populations built up?

We ran an N-body simulation of the evolution of a barred disk galaxy similar in size to the Milky Way. We initially set up an isolated disk galaxy which consists of stellar disks, with no bulge component, in a static dark matter halo potential \citep{rk12,gkc12a}, using our original Tree N-body code, {\tt GCD+} \citep{kg03a,kogbc13}. In the initial condition, we set a smaller thick disk particles and a larger thin disk particles. We first assumed the thin disk has a constant scale-height, $z_d$, independent of the radii, "constant $z_d$ model". We then assigned the metallicity of the thin disk particles to follow the radial metallicity gradient of ${\rm [Fe/H]}=0.5-0.08\times~R$ with a dispersion of 0.05~dex with no vertical metallicity gradient at $t=0$. Left two panels of Fig.~\ref{fig:zmet} show the initial (at $t=0$) and final ($t=8$) vertical metallicity distribution for the thin disk component of this model (constant $z_d$ model). Interestingly, after 8~Gyr of evolution, the vertical metallicity gradient becomes positive, which is inconsistent with the negative vertical metallicity gradients observed in the mono-age population of the thin disk as mentioned above. However, this is a natural outcome of radial mixing, because when the stars in the inner region with higher metallicity migrate outwards, they tend to end up at a higher height in the outer region of the disk, due to their higher initial vertical action.

In \citet{kggchb17} we presented that a flaring thin disk is one possibility to remedy this issue. In our N-body simulation we include a 2nd thin disk component whose vertical scale height increases with radius, "flaring model". This flaring thin disk component is representative of a mono-age population that is born in a flaring star-forming region. We assigned the metallicity following the same radial metallicity gradient as our constant $z_d$ model to the flaring disk particles. Right two panels of Fig.~\ref{fig:zmet} show the initial (at $t=0$) and final ($t=8$) vertical metallicity distribution for the flaring thin disk component of this model.  In this case, the metal poor stars formed in the outer disk become dominant at high vertical height at every radii, which can drive a negative vertical metallicity gradient. Therefore, if mono-age populations of the Milky Way thin disk stars are formed within a flaring star forming disk, each mono-age thin disk population can have a negative vertical metallicity gradient. Then, older population has a steeper negative vertical metallicity gradient, because they have more time to experience more radial mixing. This can explain the steeper negative vertical metallicity gradient observed in the older population of the thin disk \citep{cklcsc17}. 

The flaring younger thin disk population helps to explain the observed abundance distribution of thin and thick disk populations. Analysing a cosmological simulation result, \citet{rck14} found that the old compact thick disk population and the flaring younger thinner disk population leads to a negative radial metallicity gradient at the disk plane, and a positive radial metallicity gradient at the high vertical height, $2<|z|<3$~kpc, which is consistent with the observed radial metallicity gradients at different vertical heights in the Milky Way \citep[e.g.][]{ccz12}. \citet{rck14} predicted that if this is true, we should see a negative radial [$\alpha$/Fe] gradient and a negative radial age gradient at a high vertical height \citep[see also][]{mmssdjs15,mpgbsb16}. This prediction has been confirmed by several observational studies \citep[e.g.][]{acsrg14,mmnfr16}. 

\section{Metallicity Gradients of the Thick disk Progenitor}

Spectroscopic surveys of the Galactic disk stars found that the chemically-defined thick disk shows no radial metallicity gradient, but a clear negative vertical metallicity gradient \citep[e.g.][]{mhrbdv14}. This indicates that the chemical properties of the thick disk are well mixed radially, but maintain a negative vertical gradient. Combining an idealised N-body disk model with a MCMC chemical painting technique, in \citet{kalbccgghh17} we explored what metallicity distribution for the thick disk progenitor was needed, to explain the current thick disk metallicity distribution observed by the APOGEE survey \citep{hbhnb15}. 

 Similar to Sec.~\ref{sec:vZflaringthin}, we ran an N-body simulation with thin and thick disks for 8 Gyr. The main assumptions in our study is that thick disk formation was completed a long time (8~Gyr) ago and the thick disk stars experienced effective radial mixing for a long time since then \citep[e.g.][]{bsgkh12}. The simulation mimics the evolution history of the Galactic disk after the thick disk formed. At this time, we set two components for the thick disks, a thicker, smaller one (thick1) and a thinner, larger one (thick2), to mimic the inside-out and upside-down formation of the thick disk \citep[e.g.][]{bkmg06}. We study what radial and vertical metallicity gradients these two thick disk components should originally have had by comparing with the APOGEE observations of metallicity distribution function (MDF) at different Galactocentric radii and height. We run MCMC using  {\tt emcee} \citep{fmhlg13} with 8 parameters, i.e. two different sets of the initial central metallicity, [Fe/H]$_0$, radial metallicity gradient, (d[Fe/H]/dR)$_0$, vertical metallicity gradient, (d[Fe/H]/dz)$_0$, and metallicity dispersion, $\sigma_{\rm [Fe/H],0}$, at $t=0$ for thick1 and thick2 disks, to explore what parameter set would best match with the APOGEE thick disk MDFs at $t=8$ Gyr snapshot. 
 
 We found that thick1 disk prefers to be high metallicity and positive radial metallicity gradient. On the other hand, thick2 prefers to be metal rich and a negative vertical metallicity gradient. Because we set thick1 is more massive, overall the thick disk is required to have a positive radial metallicity gradient. An overall negative radial metallicity gradient for the thick disk progenitor is rejected. This is because radial mixing in a disk with an initially negative radial metallicity gradient leads to a positive vertical metallicity gradient as demonstrated in Sec.~\ref{sec:vZflaringthin}, unless they have a strong flaring disk initially. It should be difficult to have a thick disk flaring strong enough to cause a negative vertical metallicity gradient with a negative radial metallicity, because the thick disk progenitor is likely already quite kinematically hot even in the inner region, when they born \citep{bkgf04b}. However, this could be an interesting question to be explored.

  Our result of metal poor thick1 and metal rich thick2 components support a scenario in which the thick disk formed in an inside-out and upside-down fashion, and a younger, thinner and larger disk (like thick2) is more metal rich. Then, inside-out formation and age-metallicity relation produced an overall positive metallicity gradient when their formation was completed \citep[see also][]{rspm17,mcm14}. If we could measure the age of the old stars accurately enough, we would expect that the older thick disk would be smaller and more metal poor, and each generation of the thick disk would exhibit a more negative radial metallicity gradient than the overall gradient of the thick disk. It would be challenging to divide the sample of thick disk stars into age bins fine enough to test this scenario. Hence, this would be a key challenge for future work on Galactic archaeology \citep[e.g.][]{mcmdfgjk17}. 
 

\end{document}